\documentclass[]{aa}
\usepackage{txfonts}
\usepackage{natbib}
\usepackage{graphicx}
\usepackage{aalongtable}
\bibpunct{(}{)}{;}{a}{}{,}
\begin{document}
\title{A lack of close binaries among hot horizontal branch stars in globular
clusters \thanks{Based on observations with the ESO Very Large Telescope
at Paranal Observatory, Chile (proposal ID 69.D-0682)}}
\subtitle{M\,80 and NGC\,5986}

\author{
C. Moni Bidin \inst{1}
\and
S. Moehler \inst{2}
\and
G. Piotto \inst{3}
\and
Y. Momany \inst{4}
\and
A. Recio-Blanco \inst{5}
}

\institute{
Departamento de Astronom\'{i}a, Universidad de Concepci\'{o}n,
Casilla 160-C, Concepci\'{o}n, Chile
\and
European Southern Observatory, Karl-Schwarzschild-Str. 2,
85748 Garching, Germany
\and
Dipartimento di Astronomia, Universit\`{a} di Padova,
Vicolo dell'osservatorio 3, I-35122 Padova, Italy
\and
INAF-Osservatorio Astronomico di Padova,
Vicolo dell'osservatorio 2, I-35122 Padova, Italy
\and
Laboratoire Cassiop\'{e}e UMR 6202,
Universit\'{e} de Nice Sophia-Antipolis, CNRS,
Observatoire de la C\^{o}te d'Azur
}
\date{Received / Accepted }


\abstract {Recent investigations have revealed a surprising lack of close
binaries among extreme horizontal branch (EHB) stars in the globular
cluster NGC\,6752, at variance with the analogous sdB field stars.
Another puzzling result concerns the derived
spectroscopic masses for some EHB stars.}
{The present paper extends our study of NGC\,6752 to
M\,80 and NGC\,5986,
to establish whether the unexpected properties of EHB stars in NGC\,6752
are also present in other clusters.} {Twenty-one horizontal
branch stars (out of which 5 EHBs) in NGC\,5986 and 31 in M\,80
(11 EHBs) were observed during four consecutive nights. We measured
radial velocity variations and evaluated statistical and
systematic errors.  Temperatures, gravities, and helium abundances
were also measured.} {By means of a statistical analysis of the
observed radial velocity variations, we detected one EHB close binary
candidate per cluster. In M\,80, the best estimate of the close binary EHB
fraction is $f$=12\%, and even the lowest estimate of the binary fraction
among field sdB stars can be ruled out within a 90\% confidence level.
Because of the small observed sample, no strong conclusions can be drawn
on the close EHB binary fraction for NGC\,5986, although our
best estimate is rather low ($f$=25\%).\\
For the discrepancy in spectroscopic derived masses with
theoretical models observed in NGC\,6752, our analysis of M\,80 EHB stars
shows a similar trend. For the first time, we report a
clear trend in surface helium abundance with temperature,
although the trend for the hottest stars is still unclear.}
{Our results show that the deficiency of close binaries among
EHB stars is now confirmed in two, and possibly three, globular
clusters. This feature is therefore not a peculiarity of NGC\,6752.
Our analysis also proves that the strangely high spectroscopic masses
among EHB stars are now confirmed in at least a second cluster. Our results
confirm that $f$ could be a function of the age of the sdB star
population, but we find that recent models
have some problem reproducing all observations.}

\keywords{ stars: horizontal branch -- binaries: close
-- binaries: spectroscopic -- stars: fundamental parameters
-- globular cluster: individual: \object{M\,80}, \object{NGC\,5986} }

\authorrunning{}
\maketitle

\section{Introduction}
\label{capintro}

Horizontal branch (HB) stars in Galactic globular clusters
are old stars of low initial mass
(0.7-0.9 M$_\odot$) which, after the exhaustion of hydrogen
in the stellar core and the ascension along the red giant branch,
eventually ignited helium core burning \citep{Hoyle55,Faulkner66}.
The most puzzling feature of these stars is surely the large variety
of HB mophologies in globular clusters
\citep[see for example][]{Piotto02}, which is only partly explained by
differences in metallicity \citep[the so-called "second parameter
problem",][]{Sandage67,VanDenBergh67}. In this context, the foremost
problem is the presence of extreme horizontal branch (EHB) stars
at the faint hotter end of HBs (T$_\mathrm{eff}\geq$20\,000~K),
even in high metallicity clusters like \object{NGC\,6388}
and \object{NGC\,6441} \citep{Rich97}.
EHB stars are identified as hot He-core burning stars with
an external envelope too thin to sustain hydrogen shell burning,
and after He exhaustion in the core they are expected to evolve
directly to the white dwarf cooling sequence without ascending the
asymptotic giant branch \citep[AGB manqu\'{e} stars,][]{Greggio90}.
EHB stars have been extensively observed and also studied in the Galactic
field, identified as the so-called subdwarf B-type (sdB) stars
\citep{Greenstein71,Caloi72,Heber86},
although this is a spectroscopic
classification without direct link to the stellar evolutionary stage.
Given the intrinsic faintness of these objects, they are still 
spectroscopically poorly observed in globular clusters, and many open
problems lack full comprehension \citep[see][for recent reviews]{Moni07b,Catelan05}.

One of the most evident features of HB stars is the onset of
atmospheric diffusion for temperatures
$T_\mathrm{eff}\geq$11\,000-12\,000~K. This causes changes in
their photometric properties \citep[][]{Grundahl99},
deficiency of helium because of gravitational settling,
and strong (solar to super-solar levels) enrichment of heavy metals
(\citealt{Glaspey85},
\object{NGC6397}; \citealt{Glaspey89,Moehler00}, \object{NGC6752};
\citealt{Behr99,Moehler03}, \object{M13}; \citealt{Behr00},
\object{M15}; \citealt{Fabbian05}, \object{NGC1904};
\citealt{Pace06}, \object{NGC2808}).
Precise calculations confirmed that diffusion should be at
work in the atmospheres of these stars and can account for observed
anomalies \citep{Michaud83}. \citet{Michaud08} recently
confirmed the role of atomic diffusion in the observed abundance
anomalies using new stellar evolution models.
Spectroscopically determined
surface gravities are systematically lower than predictions 
for stars in this temperature range \citep{Moehler95}, a problem
only partially explained by abundance anomalies \citep{Moehler01}.
\citet{Vink02} point out that neglecting the presence of stellar
wind can cause measured surface gravities to be erroneously low,
and an enhanced stellar wind is actually the explanation that
\citet[][hereafter M07]{Moni07a} proposed for some bright stars
showing erroneously low masses.
It is worth noting, however, that optical and UV observations do
not support high mass loss rates for field EHB stars in general, with the
exception of few relatively luminous objects
\citep{Maxted01,Lisker05}.
\defcitealias{Moni07a}{M07} \citet{Momany02,Momany04}
found that at temperatures hotter than 23\,000~K HB stars deviate
again from canonical tracks in the color-magnitude diagram, and
they proposed a new onset of diffusion as explanation of this
feature. The low helium abundances found by \citet{Moehler00}
and \citetalias{Moni07a}
on some hot stars seem to confirm this hypothesis, but the pattern
of abundance with temperature is unclear.

In canonical models, the mass of the He-burning core is
approximatively the same for all HB stars, equal to the minimum
required for core helium flash
\citep[$\approx 0.5$M$_{\odot}$,][]{Schwarzschild62},
but the envelope mass decreases for
higher temperature. The extremely hot EHB stars retain just a very
thin inert hydrogen envelope \citep[$\leq$0.02M$_{\odot}$,][]{Heber86},
and must have suffered an extreme mass loss during their evolution.
Many single-star evolutionary channels have been invoked to explain
EHB star formation in globular clusters, including interactions with a close planet
\citep[][see also \citealt{Silvotti07}]{Soker98}, He mixing driven by
internal rotation \citep{Sweigart79,Sweigart97} or by stellar encounters
\citep{Suda07}, dredge-up induced by H-shell instabilities
(\citealt{vonRudloff88}, but see also \citealt{Denissenkov03}), close
encounters with a central, intermediate-mass black hole
\citep{Miocchi07}, and a sub-population of stars with high helium
abundance \citep[e.g.,][]{DAntona05}.
The discovery of multiple main sequences in \object{$\omega$ Cen}
\citep{Bedin04} and in NGC\,2808 \citep{Piotto07} reinforced the idea that
in some clusters there might be a fraction of stars super-He rich, up to
Y$\sim$0.40 \citep{Norris04,Piotto05,DAntona05,Lee05}.
Nevertheless binary models, in which
sdB stars form through dynamical interactions within binary
systems, have been very successful in reproducing observations
\citep{Han02,Han03,Han07}, and are actually the most preferred scenario
for field sdB star formation. Indeed, many surveys have shown the existence
of a large population of sdB binaries
\citep{Ferguson84,Allard94,Ulla98,Aznar01,Maxted01,Williams01,Reed04,Napiwotzki04}.
Among them, close systems with periods shorter than 10 days play a major
role \citep{Moran99,Saffer98,Heber02,MoralesRueda03}.
The close binary fraction among field sdB stars is certainly high but
still ill-determined, ranging from 70\% \citep{Maxted01} to 40-45\%
\citep{Napiwotzki04}. In this context, it came as a great surprise that
first surveys in globular clusters revealed a lack of close binary systems among the
EHB stars \citep[][hereafter Paper~I]{Moni06a}.
\defcitealias{Moni06a}{Paper~I}
Recently \citet{Moni08} showed that the best estimate of the
close binary fraction among
EHB stars in \object{NGC\,6752} is only $f$=4\%.
On the basis of theoretical and observational results available
in the literature, they suggested the presence of a $f$-age relation.
\citet{Han08} supported
this hypothesis with detailed theoretical calculations, showing that
the binary scenario naturally predicts a steep decrease of close binary
fraction with increasing age of the sdB population.
This seems a further success of Han's models,
but the general lack of observational data in globular clusters and the uncertainties
on the predicted $f$ values (due to uncertainties on model parameters) still
requires caution.

Observations of globular clusters other than \object{NGC\,6752} are needed
to verify that the low fraction of close EHB systems is
not a peculiar feature of this cluster. Also other types of
binary systems must be included in future searches. In fact, both wide binaries
\citep{Reed04,MoralesRueda06} and systems with very low-mass
secondaries \citep{Menzies86} are known to exist among field sdBs
(although they are a minor population there) and it has been shown that
they can provide an additional channel for the formation of sdB stars.

In this paper we present the results of a binary search in
two additional globular clusters,
\object{NGC\,5986} and \object{M\,80}. A first analysis of data was
presented in \citet{Moni06b}. Here we refine that preliminary
overview with the correction of systematic effects and a
detailed error analysis, and we extensively use statistical
calculations to better clarify the significance of our results.
We also present results about atmospheric
parameters and masses for our target stars.


\section{Observations and data reduction}
\label{capdata}

We selected 21 HB stars in \object{NGC\,5986} and 31 in
\object{M\,80}, spanning a wide range in effective temperature.
Targets in \object{M\,80} were divided in two fields, named M\,80a
(17 stars) and M\,80b (15 stars), with different slit configurations
for multi-object spectroscopy. One star in \object{M\,80} (\#14327) was
accidentally observed in both fields (as star 1a and 12b respectively).
The positions of the observed stars along the HB of their parent cluster
are shown in Fig.~\ref{cmd}, while astrometric and photometric
data from \citet{Momany03} and \citet{Momany04} are presented in
Table~\ref{tabdata}.

\begin{figure}
\begin{center}
\resizebox{\hsize}{!}{\includegraphics{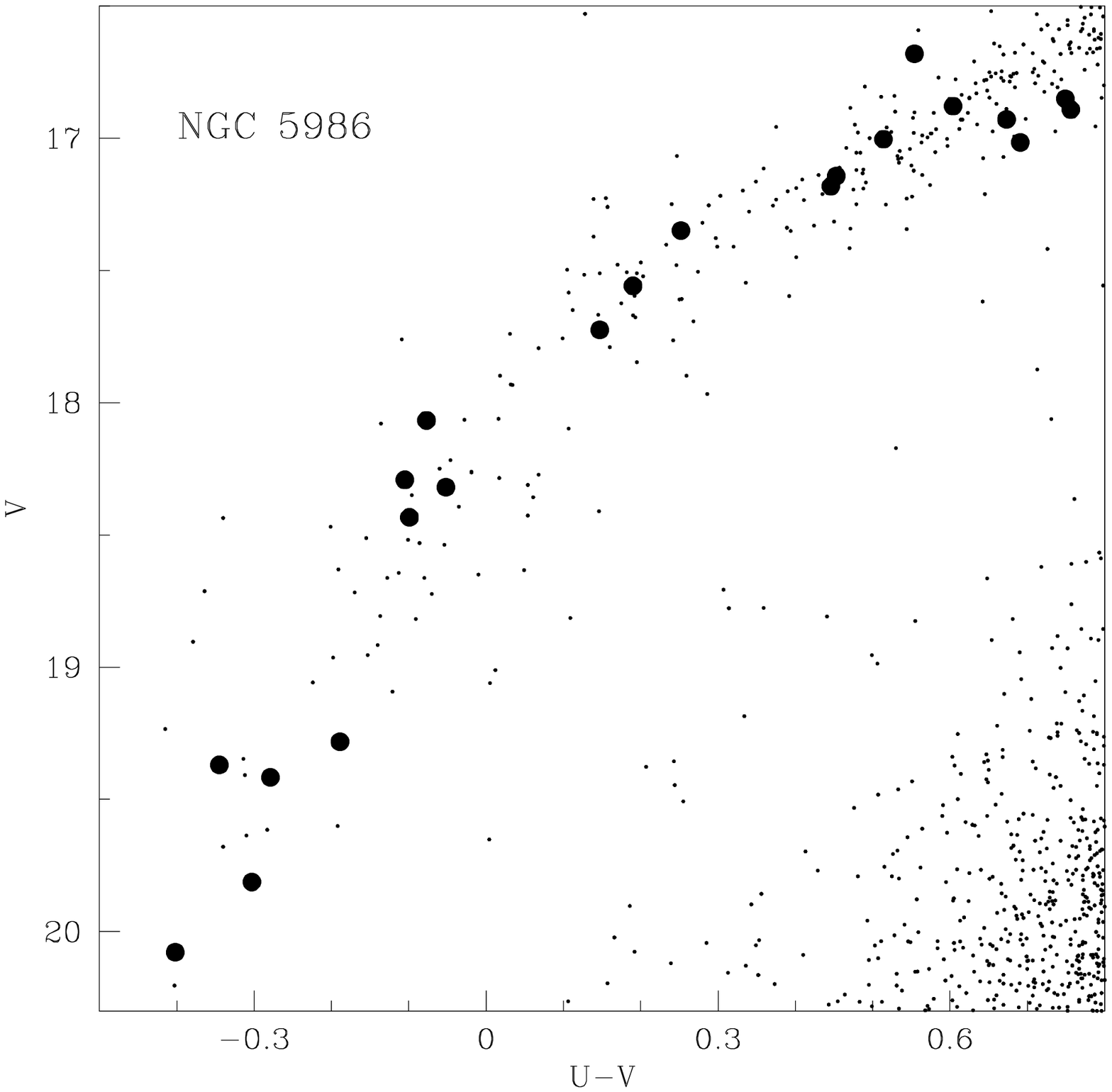}}
\end{center}
\end{figure}
\begin{figure}
\begin{center}
\resizebox{\hsize}{!}{\includegraphics{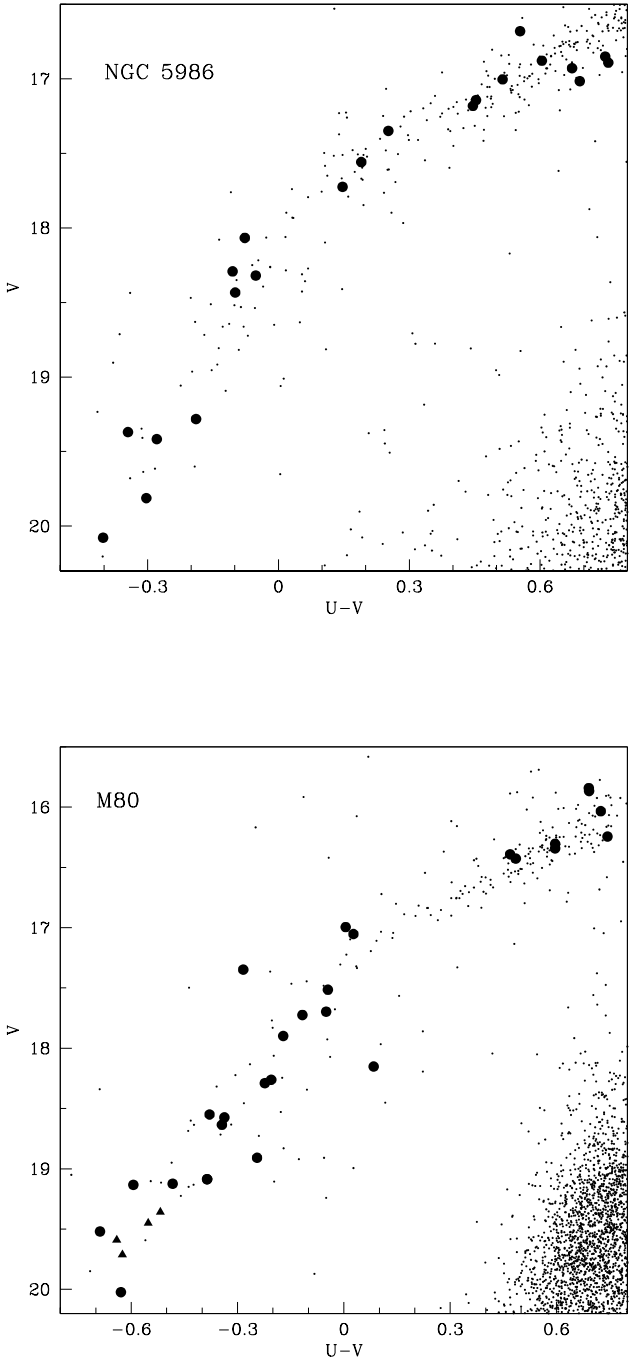}}
\caption{Color-magnitude diagram of NGC\,5986 ({\it upper panel})
and M\,80 ({\it lower panel}), with targets highlighted.
{Stars for which we derive anomalously high masses are
plotted as full triangles.}
Photometric data are from \citet{Momany03} for NGC\,5986 and
\citet{Momany04} for M\,80.}
\label{cmd}
\end{center}
\end{figure}

The spectra were collected during four nights of observations (June
11th to 14th, 2002) at the VLT-UT4 telescope equipped with the
spectrograph FORS2 in MXU mode. We employed the grism 1400V with
0\farcs5- wide slits, and the resulting resolution was 1.2~\AA.
The 2400\,s exposures were always acquired in pairs and subsequently
summed, with the exception of one single additional spectrum of
\object{NGC\,5986} targets during the second night. We finally
decided to exclude it from analysis because of its too low S/N,
which caused unreliable measurements. The bias, flat, and lamp exposures
were acquired during daytime. Just
before each pair of exposures, a slit image (without grism) was taken,
which was used to correct the spectroscopic data for instrumental effects.
The spectral range varied from star to star because of
different positions of the slit in the mask, but the $\mathrm{H_{\beta}}$
line was always inside the spectral range.

\begin{table}[t]
\begin{center}
\caption{UT at the start of the first of each pair of exposures.}
\label{tabtime}
\begin{tabular}{c| c c c c}
\hline
\hline
Field & \multicolumn{4}{|c}{night} \\
 & 12 & 13 & 14 & 15 \\
\hline
NGC\,5986 & 1:18 & 23:19 & 3:00 & 3:00 \\
& 3:23 &  &  &  \\
\hline
M\,80a & -- & 6:20 & 23:27 & 4:30 \\
&  &  & 0:53 &  \\
\hline
M\,80b & -- & -- & 4:31 & 23:32 \\
&  &  &  & 0:57 \\
\hline
\end{tabular}
\end{center}
\end{table}

During each night, we successfully collected at least one pair of medium-resolution
spectra for \object{NGC\,5986}, and the resulting temporal sampling was
very good.
Unfortunately, because of strong winds from north, observations of \object{M\,80}
could not be carried out during the entire first night and partially the second one.
The starting time of the pairs of exposures is given in Table~\ref{tabtime}.

During the same run we took two 1350\,s exposures of each target
with grism 600B, for a resulting resolution of 3\AA, to measure
atmospheric parameters.
Spectra were trimmed at 3600~\AA\ on the
blue side, because of the lack of instrumental response and atmospheric
transmission. All Balmer lines from H$_{\beta}$ to H10 were always
present in these spectra.

Data reduction was performed with standard MIDAS\footnote{ ESO-MIDAS
is the acronym for the European Southern Observatory Munich Image
Data Analysis System which is developed and maintained by the
European Southern Observatory
(http://www.eso.org/projects/esomidas/)} procedures. All slitlets
were trimmed from the multi-object frames and reduced independently.
The wavelength calibration (wlc) was performed using He and HgCd lamp
exposures, fitting a $3^\mathrm{rd}$ order polynomial to the dispersion
relation for both grisms. The mean rms of this fit for the
medium-resolution spectra was 2.46$\cdot$10$^{-2}$\AA. All two-dimensional
spectra were corrected for curvature along the spatial axis tracing them
with a specific MIDAS routine, as described in \citet{Moehler06}, and
then extracted with an optimum extraction algorithm \citep{Horne86}.
Sometimes (usually for brighter stars) this procedure failed, producing
noisy spectra with irregular continuum, and we opted for a simple sum
algorithm in these cases. 1400V spectra were rebinned to constant steps of
0.25~\AA /pix, and continuum-normalized.
Most spectra of \object{M\,80} stars showed a tiny but clear interstellar
emission line in the core of H$_{\beta}$. During data reduction we
gave particular attention to the proper removal of this feature,
which could spoil the RV variation measurements. 600B spectra
were rebinned to a larger constant step (0.4~\AA /pix), corrected for
atmospheric extinction with the coefficients for La Silla observatory
\citep{Tug77}, and flux-calibrated.
The response curve was obtained separately for
each night, through observations of standard stars \object{EG274} and
\object{LTT3218} with the flux table of \citet{Hamuy94}. Finally, on
600B spectra we fitted a Gaussian profile to the core of all Balmer
lines from H$_{\beta}$ to H$_{9}$, excluding H$_\epsilon$ due to
blending with the \ion{Ca}{ii}~H line, and we used the resulting
average radial velocity to shift the observed spectra to laboratory wavelengths.

\begin{table}[h!]
\begin{center}
\caption{Photometric data of program stars. Columns 1: slit number. Columns 2-6: IDs, coordinates
and photometric data from \citet{Momany03} and \citet{Momany04}.}
\label{tabdata}
\small
\begin{tabular}{ c c c c c r }
\hline \hline
slit&ID&RA (J2000)&DEC (J2000)&V&($U-V$)\\
&& hh:mm:ss&$^{\circ}$: ' : ''&&\\
\hline
\multicolumn{6}{c}{NGC\,5986} \\
\hline
1&17512&15:46:06.215&$-$37:50:49.84&18.320&$-$0.052\\
2&17604&15:46:03.177&$-$37:50:41.41&16.892&0.756\\
3&17691&15:46:08.614&$-$37:50:29.56&19.812&$-$0.303\\
4&18077&15:46:05.939&$-$37:49:59.20&17.016&0.691\\
5&18240&15:46:07.543&$-$37:49:47.01&16.851&0.749\\
6&2103&15:46:10.144&$-$37:48:31.16&17.559&0.190\\
7&3192&15:46:11.382&$-$37:48:06.34&18.067&$-$0.077\\
8&3560&15:46:12.748&$-$37:47:58.32&18.434&$-$0.099\\
9&4175&15:46:16.056&$-$37:47:44.62&19.417&$-$0.279\\
10&4930&15:46:18.214&$-$37:47:29.39&17.143&0.453\\
11&5558&15:46:11.971&$-$37:47:21.61&17.349&0.252\\
12&6102&15:46:08.134&$-$37:47:13.91&17.725&0.147\\
13&7008&15:46:13.634&$-$37:46:54.66&16.879&0.604\\
14&7430&15:46:11.091&$-$37:46:48.42&17.004&0.514\\
15&8131&15:46:11.505&$-$37:46:34.93&19.370&$-$0.345\\
16&9049&15:46:04.640&$-$37:46:21.30&18.292&$-$0.105\\
17&9250&15:46:17.684&$-$37:46:09.79&16.929&0.673\\
18&10390&15:46:09.599&$-$37:45:51.45&16.681&0.554\\
19&11215&15:46:11.234&$-$37:45:30.85&17.182&0.446\\
20&11571&15:46:16.004&$-$37:45:17.74&20.078&$-$0.402\\
21&12099&15:46:09.689&$-$37:45:05.36&19.282&$-$0.189\\
\hline
\multicolumn{6}{c}{M\,80} \\
\hline
1a&14327&16:17:11.476&$-$22:59:23.30&19.086&$-$0.386\\
2a&14786&16:17:10.954&$-$22:59:06.21&16.304&0.596\\
3a&14985&16:17:08.723&$-$22:58:59.74&18.290&$-$0.223\\
4a&15200&16:17:06.303&$-$22:58:53.05&18.152&0.084\\
5a&16389&16:17:12.046&$-$22:58:05.56&18.549&$-$0.379\\
6a&16163&16:17:05.336&$-$22:58:18.14&17.724&$-$0.117\\
7a&17173&16:17:10.234&$-$22:57:37.98&16.244&0.744\\
8a&17114&16:17:07.499&$-$22:57:42.36&17.515&$-$0.045\\
9a&16707&16:16:56.026&$-$22:58:04.77&20.023&$-$0.629\\
10a&17737&16:17:03.158&$-$22:57:21.75&19.359&$-$0.518\\
11a&18516&16:17:07.675&$-$22:56:45.43&18.635&$-$0.344\\
12a&18110&16:17:00.961&$-$22:57:07.79&17.053&0.027\\
13a&18992&16:17:08.007&$-$22:56:18.24&16.391&0.469\\
14a&19040&16:17:06.095&$-$22:56:16.68&18.574&$-$0.337\\
15a&18391&16:16:54.498&$-$22:56:59.78&15.842&0.691\\
16a&19191&16:17:03.050&$-$22:56:07.66&19.124&$-$0.483\\
17a&19127&16:16:55.754&$-$22:56:16.88&19.450&$-$0.552\\
1b&12304&16:16:58.233&$-$23:01:19.25&19.590&$-$0.641\\
2b&14201&16:16:57.762&$-$22:59:37.14&17.898&$-$0.171\\
3b&13787&16:16:59.111&$-$22:59:53.45&18.261&$-$0.205\\
4b&12663&16:17:01.896&$-$23:00:48.22&19.133&$-$0.59\\
5b&12767&16:17:03.678&$-$23:00:40.16&19.712&$-$0.625\\
6b&13839&16:17:03.211&$-$22:59:48.33&16.342&0.596\\
7b&14022&16:17:05.686&$-$22:59:38.95&18.909&$-$0.245\\
8b&14387&16:17:06.469&$-$22:59:24.31&17.697&$-$0.050\\
9b&12526&16:17:10.297&$-$23:00:52.02&19.520&$-$0.688\\
10b&13179&16:17:10.249&$-$23:00:12.68&17.348&$-$0.284\\
11b&15682&16:17:08.241&$-$22:58:34.56&16.995&0.005\\
12b&14327&16:17:11.476&$-$22:59:23.30&19.086&$-$0.386\\
13b&15183&16:17:12.947&$-$22:58:49.48&15.866&0.692\\
14b&14813&16:17:14.702&$-$22:59:02.51&16.427&0.485\\
15b&15470&16:17:15.921&$-$22:58:37.19&16.034&0.725\\
\hline
\end{tabular}
\end{center}
\end{table}


\section{Measurements}
\label{capmeasure}

\subsection{Atmospheric parameters and masses}
\label{capmeasureparam}

To derive atmospheric
parameters by means of Balmer and helium lines fitting, it is important
to know whether the stellar atmosphere is affected by diffusion
processes. In fact, the profile of lines under study can be influenced by 
helium depletion and metal enrichment.
\citet{Moehler99,Moehler00,Moehler03} showed that strong
\ion{Fe}{ii} lines in the region 4450-4600~\AA\ are detectable
even at low resolution when diffusion starts up at about 12\,000~K
\citepalias[see for example Fig.~3 in][]{Moni07a}.
Therefore, spectra showing evidence of iron lines
{\em or} of being hotter than 14\,000~K (as deduced from their position
in the color-magnitude diagram) were fit with metal-rich
([M/H] $+0.5$) model spectra, whereas we adopted metal-poor models
([M/H] = $-1.5$) for all other stars. In few cases we relied
also on higher resolution 1400V spectra, summed altogether for a
higher S/N, in search for evidence of atmospheric diffusion.
We kept the helium abundance fixed at solar value
($\log{\frac{\mathrm{N_{He}}}{\mathrm{N_H}}}=-$1.00)
for cool stars (T$_\mathrm{eff}\lesssim$11\,000\,K), as the helium lines in
their spectra are rather weak.
During the fitting we verified that helium lines predicted for
these targets agreed with the observed ones.

We computed model atmospheres using ATLAS9 \citep{Kurucz93} and
used Lemke's version\footnote{For a description see
http://a400.sternwarte.uni-erlangen.de/$\sim$ai26/linfit/linfor.html}
of the LINFOR program (developed originally by Holweger, Steffen,
and Steenbock at Kiel University) to compute a grid of theoretical
spectra that include the Balmer lines H$_\alpha$ to H$_{22}$,
\ion{He}{i} (4026~\AA, 4388~\AA, 4471~\AA, 4921~\AA), and
\ion{He}{ii} lines (4542~\AA\ and 4686~\AA). The grid covered the
range 7\,000~K~$\leq$~T$_\mathrm{eff}$~$\leq$~35\,000~K,
2.5~$\leq$~$\log{g}$~$\leq$~6.0,
$-3.0$~$\leq$~$\log{\frac{\mathrm{N_{He}}}{\mathrm{N_H}}}$~$\leq$~$-1.0$,
at metallicities of
[M/H]~=~$-1.5$ and $+$0.5. To establish the best fit to the observed
spectra, we used the routines developed by \citet{Bergeron92} and
\citet{Saffer94}, as modified by \citet{Napiwotzki99},
which employ a $\chi^2$ test. The $\sigma$ necessary for the
calculation of $\chi^2$ is estimated from the noise in the continuum
regions of the spectra. The fitting program normalizes model
{\em and} observed spectra using the same points for the continuum
definition. H$_\epsilon$ was excluded to avoid the blended
\ion{Ca}{ii}~H line. The errors in each fitting procedure were
derived from the $\chi^2$ of the fit itself \citep[see][for more
details]{Moehler99}, under the assumption that the only error source
is the statistical noise. However, Napiwotzki (priv. comm.) noted that
the routine underestimates this statistical error by a factor of 2-4.
In addition, errors in the normalization of the spectrum, imperfections
of flat field/sky background correction, etc. may produce systematic
errors, which are not well represented by the error obtained from the
fit routine.

For each star, we measured atmospheric parameters in the two
600B spectra separately, and the final results are the weighted mean
of the two measurements. Errors were multiplied by $\sqrt{3}$
because the fitting procedure assumes each pixel as independent of
the others, but when rebinning we oversampled the spectra by a factor
of three with respect to the dispersion. Stars 1a and 12b in
\object{M\,80}, which are actually the same object, were studied as if
they were two different targets, so we could compare the results
as an indication of their quality. Results are in excellent
agreement (see Table~\ref{tabres}), despite the target being among
the faintest
ones, showing that the reported internal errors are
probably realistic. The parameters
from the two distinct measurements are so similar that we found
no need to weight them altogether, hence in our analysis we will
simply omit star 12b in order to assure statistical uniformity to our
sample. Here, the choice of which star to exclude is
quite irrelevant, and we opted to keep star 1a for continuity
with RV variation analysis, where the two slitlets are not equivalent
because of the better time coverage of field M\,80a (see discussion
in \S \ref{capmeasurervvar}).

Masses were calculated from the previously measured atmospheric
parameters, through the equation:
\begin{equation}
\log{\frac{M}{M_{\sun}}}= \log{\frac{g}{g_{\sun}}} - 4\cdot
\log{\frac{T}{T_{\sun}}} - \frac{M_\mathrm{V} + BC - 4.74}{2.5},
\label{eq1}
\end{equation}
obtained from basic relations. We adopted the standard values
T$_{\sun}$=5777~K and $\log{g_{\sun}}$=4.4377.
The bolometric correction was derived from effective temperature
through the empirical calibration of \citet{Flower96}.
We adopted an apparent distance modulus (m$-$M)$_{V}$=15.96 for
\object{NGC\,5986} and (m$-$M)$_{V}$=15.56 for \object{M\,80}
\citep[][February 2003 Web version]{Harris96}. Errors on mass estimates
were derived from propagation of errors, assuming an uncertainty of 0.1
on photometric quantities (distance modulus, magnitude and BC).

Our results are presented in Table~\ref{tabres}.
Despite the fact that instruments, observing nights,
and measurement procedures
were the same as in \citetalias{Moni07a}, the errors are much higher
here, This is easily explained by the faintness of the targets in
these two clusters with respect to NGC6752.

\begin{figure}
\begin{center}
\resizebox{\hsize}{!}{\includegraphics{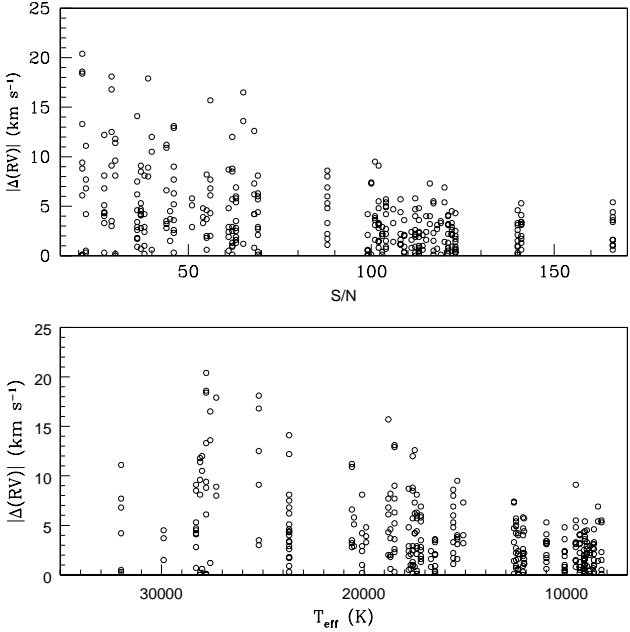}}
\caption{{\it Upper panel}: absolute value of measured RV variations
of all stars plotted against the spectral S/N. {\it Lower panel}: same
plot but as a function of stellar effective temperature.}
\label{sn}
\end{center}
\end{figure}

\subsection{Radial velocity variations}
\label{capmeasurervvar}

Radial velocity (RV) variations were measured by means of the
cross-correlation (CC) technique \citep{Tonry79}. The targets are
fainter by 1-2 magnitudes with respect to \citetalias{Moni06a}, and
spectra were much noisier despite the longer exposure times. As a
consequence, the measurement procedure presents some small difference
with respect to previous work, and final errors are higher, 
especially for hot stars.

We cross-correlated each
spectrum with all the others of the same star using the
IRAF\footnote{IRAF is distributed by the National
Optical Astronomy Observatories, which are operated by the
Association of Universities for Research in Astronomy, Inc., under
cooperative agreement with the National Science Foundation.} task
{\it fxcor}. Thus, we measured 10 RV variations for each target in
NGC\,5986 (5 spectra per star), 6 in M\,80a (4 spectra), and only 3
in M\,80b (3 spectra). To define the sign of the variation, unnecessary
for our goals, the first spectrum in temporal order was always assumed
as template. We cross-correlated the single spectra of each pair
before summing them, to check that the sum was safe and no RV
variation occurred in between exposures. 
Unfortunately, single spectra of very hot stars were often of too
low S/N for a reliable CC, and they were summed {\it bona fide}.

RV variation measurements focused on H$_{\beta}$ line with full
wings. Based on our experience, the CC of such a wide line
fails to give the correct shift of the spectra if restricted to line
cores, and results tend to lower values if the adopted interval is
too narrow. On the other hand, on larger intervals, the results
converge to a fixed value, but the line wings provide progressively
less information and more noise. We performed tests on
artificially shifted spectra, looking for the best compromise, and
eventually we adopted the interval 4830-4890~\AA\ in our CC.
The CC function roughly resembles the line profile, and in
our case it took the shape of a Gaussian core with wide wings.
We performed a Gaussian fit of the central peak to determine its
center. For hotter stars, the low S/N (see Fig.~\ref{sn})
required the application
of a  Fourier filter \citep{Brault71}. The resulting
CC function lost its strong irregularities and was easier to fit,
but we always verified that RV variation after filtering did not
differ from what could be (sometimes with difficulties) deduced
fitting the unfiltered CC function

We always tried to confirm the measured RV variation cross-correlating
weaker spectral lines. On hotter stars this proved fruitless, because
too low S/N caused unreliable (often even impossible) measurements.
Results agreed with H$_{\beta}$, but within so large errors that the
confirmation was useless. On cooler stars we obtained very good CCs but
they just confirmed measurements on H$_{\beta}$, without additional
information. Hence, we will not analyze results obtained with lines
other than H$_{\beta}$.

\begin{figure}
\begin{center}
\resizebox{\hsize}{!}{\includegraphics{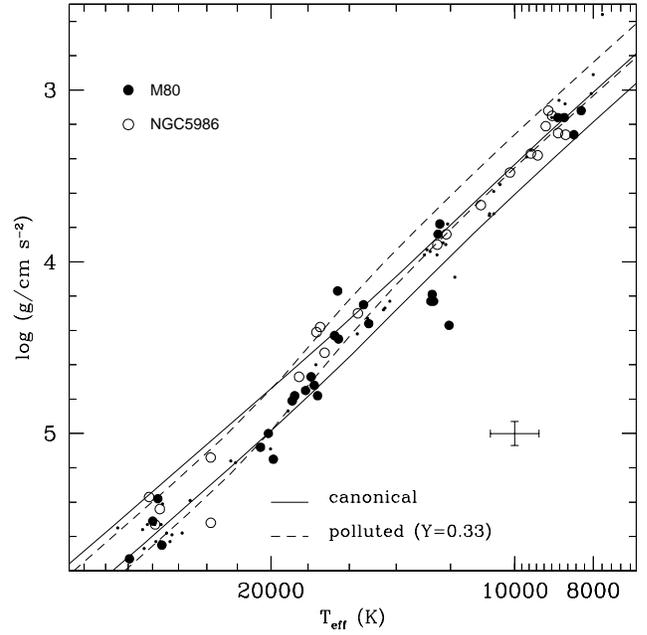}}
\caption{T$_\mathrm{eff}$-log(g) plot of observed stars.
(full points: \object{NGC\,5986}, open points: \object{M\,80}).
Errorbars on single stars are omitted for clarity, but the
errorbar drawn in lower right corresponds to typical values
$\frac{\sigma(T_\mathrm{eff})}{T_\mathrm{eff}}$=0.03,
$\sigma$(log(g))=0.07 dex. Small points indicate results on
NGC\,6752 from \citetalias{Moni07a}.
Zero-age (ZAHB) and terminal-age (TAHB) horizontal branch
theoretical tracks are also indicated, for [M/H]=-1.5 and
canonical models (Y=0.24) and polluted ones (Y=0.33)
\citep[see][for details]{Moehler03}.}
\label{tgplot}
\end{center}
\end{figure}

\subsubsection{Corrections on radial velocity variations}
\label{capmeasurecorr}

RV variations were corrected as in \citetalias{Moni06a}.
The reader is referred to \citetalias{Moni06a} for an extensive
discussion. In brief, we first used the [\ion{O}{i}] $5577$~\AA\ sky emission
line as a zero-point to correct the spectral shift with respect to the arc
lamp observations. The position of the forbidden line was determined
with a Gaussian fit. After this, we corrected RV variations caused
by different positions of the stars within the slits, using the
slit images secured before each pair of exposures.

The corrections to be applied clearly correlate with the spatial
Y coordinates on the CCD \citepalias[see Fig.~5 and 7 of][]{Moni06a}.
In both steps we preferred to obtain the final corrections
from a least-square fit as a function of Y, to reduce the
additional noise added to the final results. The correction
procedure on field M\,80b was straightforward. On NGC5986 and
M\,80a we sometimes found mismatches between the derived corrections
and RV variations to be corrected, in a way very similar to
Fig.~7 of \citetalias{Moni06a} (middle panel). This indicates
movements of the mask inside its frame between the slit and
science images, as already discussed in \citetalias{Moni06a}.
On the other hand, we found negligible indication of rotation of
the masks (i.e. different slope between corrections and RV
variations when plotted against spatial coordinate).
It was recently found\footnote{see
http://www.eso.org/observing/dfo/quality/FORS2/reports/ \\
HEALTH/trend\_report\_LSS\_lambda\_c\_T\_HC.html}
that the central wavelength for FORS2 data varies with the 
temperature at the telescope focus.
This fact also could explain the observed shifts of
spectra on CCD, but we found no clear correlation between them
and focus temperature differences as obtained from frame headers.
Moreover, no shift was
observed among spectra in field M\,80a, as among many frames
studied by \citetalias{Moni06a}, despite the fact that
temperature changes occurred. We must conclude that in some cases
we observe small rigid shifts of spectra on CCD not caused by
different centering of stars in the slits, that can be
corrected because they affect simultaneously all the spectra in
the same frame \citepalias[see][for details]{Moni06a}. They cannot
be explained by variations of the temperature at telescope focus
alone, and movements of the mask within its frame must play a role,
although the two causes could act together and sum their effects.

\subsubsection{Errors on radial velocity variations}
\label{caperrors}

\begin{table}[t]
\begin{center}
\caption{"Extraction and fit" errors (in km s$^{-1}$) for each field
and different S/N range.}
\label{taberr}
\begin{tabular}{c| c c c }
\hline
\hline
S/N & \multicolumn{3}{c}{field} \\
 & NGC\,5986 & M\,80a & M\,80b \\
\hline
$\geq$80 & 1.2 & 1.2 & 1.0 \\
40--80 & 2.1 & 3.0 & 3.0 \\
$\leq$40 & 6.3 & 5.0 & 5.4 \\
\hline
\end{tabular}
\end{center}
\end{table}

The analysis of RV variations in search of binary systems
requires an accurate error analysis, because the crucial point is
to tell if variations are consistent with random measurement errors
or are an indication of binarity. As discussed later (\S
\ref{capresultsrvar}), varying estimated errors by 10\% can noticeably
change the probability of the datum, up to a factor of two.

We estimated the error associated to each RV variation as the
quadratic sum of all relevant sources. The wavelength calibration error
was deduced from the rms of the calibration procedure, and it was directly
calculated by the calibration routine. Its exact value changed from
spectrum to spectrum, but the differences were negligible in the final
quadratic sum, therefore we kept this contribution fixed to the mean
value of 1.5 km s$^{-1}$. This is in good agreement with the error
estimated in \citetalias{Moni06a} for the same instrumentation, obtained
analyzing calibrated lamp images. The error introduced by the
correction of systematic effects (\S \ref{capmeasurecorr}) were
estimated as in \citetalias{Moni06a}, from the uncertainty in the sky
line peak position, and the scatter of residuals with respect to the
least-square solutions used for the corrections. We did not find any
relevant difference with respect to \citetalias{Moni06a}, and we
adopted a fixed $\sigma_\mathrm{sky}$=1.5 km s$^{-1}$ for the first
correction uncertainty and $\sigma_\mathrm{corr}$=1.1 km s$^{-1}$ for
the second one. Wavelength calibration and sky-line correction errors were
considered twice in the quadratic sum, because in each RV variation two
spectra were involved.

Additional uncertainties were introduced by the choice of the
parameters determining the CC function fit and of the spectrum
extraction, because different extractions caused noise-induced
differences in the line profile. They
were estimated re-extracting spectra in slightly different ways
and re-fitting CC functions for all hot stars and a sample of cooler
ones. The dispersion of the differences between the measurements was
assumed as an estimate of the error. This parameter is very sensitive to
spectral noise, and we grouped stars in each cluster in three ranges
of S/N. The resulting errors are given in Table~\ref{taberr}.
In Fig.~\ref{sn} we plot the absolute value of RV variations for
all stars, as a function of their S/N and temperature, excluding the
three binary candidates discussed in \S\ref{capresultsrvar} (two EHB and
a cool target). The trend
of decreasing dispersion with increasing S/N is clear.

\subsection{Absolute radial velocities}
\label{capmeasureabsrv}

Absolute RVs were measured on 1400V spectra by means of CC with
synthetic spectra of appropriate temperature and gravity drawn from the
library of \citet{Munari05}. We verified that the template metallicity
had negligible effects on the results. Each measurement was corrected
to make the [\ion{O}{i}] $5577$~\AA\ sky line coincide with its
laboratory wavelength, then the weighted mean was
computed to derive the final absolute RV. The error resulting from
the weighting procedure was unrealistically small, and for the final
error we adopted the dispersion of the single measurements.
Results are shown in Table~\ref{tabres}.

Within errors, all stars show an absolute RV compatible with parent
cluster. We can conclude that \object{NGC\,5986} targets are most
likely cluster members. For \object{M\,80} stars we can just state
that RV does not disprove cluster membership. In fact, because of
the very low
cluster RV \citep[8.2 km s$^{-1}$,][February 2003 Web version]{Harris96},
we cannot distinguish
members from foreground Galactic disk stars, which are expected to
contaminate the field at such low Galactic latitudes
\citep[$b=19\fdg5$,][February 2003 Web version]{Harris96}.
Indeed, we have reasons to suspect that some
cool stars are main sequence foreground objects (as discussed in
\S\ref{capresultsparam}), but their absolute RVs do not noticeably
differ from the cluster value.
The RV of star \#14327 in \object{M\,80} was measured separately
in both fields, but results are identical in value and error.


\section{Results on atmospheric parameters and masses}
\label{capresultsparam}

The 600B spectrum of star \#16707 in \object{M\,80} fell between the
two FORS2 chips, and we could not measure its atmospheric parameters.
Its 1400V spectra were well inside the second chip,
because of the offset between the two grisms in the spatial direction,
and RV variation measurements proceeded normally. We deduced its effective
temperature from its color, for the only purpose of its classification
in our search for binarity (\S\ref{capbinaries}).

Our results on effective temperatures and gravities are plotted in
Fig.~\ref{tgplot}, where we compare the position of our target stars
with theoretical models.
The zero-age (ZAHB) and terminal-age (TAHB) HB theoretical tracks
for [M/H]=$-$1.5 from \citet{Moehler03} are indicated,
both for canonical models of normal
He-content (Y=0.24) and polluted ones with enhanced helium abundance
(Y=0.33). The ZAHB and TAHB define the region where models spend
99\% of their HB lifetime.

In Fig.~\ref{tgplot} four \object{M\,80} stars at
T$_\mathrm{eff}\approx$12\,000~K stand out for their too high
gravities. The fitting procedure outlined in \S\ref{capmeasureparam}
was usually problematic for these targets, although both 600B and
1400V spectra were of good quality and no problems were encountered
during reduction. Their spectra show a high quantity of strong metallic lines,
but the fit with super-solar metallicity models is not convincingly
better and, on the contrary, for two of these targets
it was too poor to permit parameter determination. For this reason,
the tabulated and plotted results throughout this paper refer to fits
with parent cluster metallicity models, although they should not be the
more appropriate. Their derived masses are very high compared to canonical
values (2.5-5 M$_{\odot}$), and temperatures are completely inconsistent
with their color: these stars can be found in Fig.~\ref{cmd} at the
reddest end of the HB ($U-V \approx$0.7), far from other HB stars with
the same temperature and redder than the coolest "normal" HB stars with
T$_\mathrm{eff}$=8\,500~K. All these results point to a mismatch between
their spectra and theoretical models, and we suspect these objects are
foreground main sequence stars. This hypothesis is further corroborated by
their metal-rich spectra,
that would be very surprising for HB cluster members much redder
than the Grundahl jump. Both temperatures and masses are roughly consistent
with this explanation, but they might not be reliable.
In particular, masses
could be overestimated by Eq. \ref{eq1}, due to a wrong distance modulus.
For one of these four stars we detect some RV variability on 1400V spectra,
as analyzed in \S\ref{capbinaries}.

Our results agree fairly well with theoretical expectations, and, within errors,
the global behavior of all the points in Fig.~\ref{tgplot} is to follow the
canonical track, although the presence of some He-enhanced stars
can not be excluded. 
One may note that the tendency toward lower gravities in the range
T$_\mathrm{eff}$=12\,000-17\,000~K, that moves the points closer to
the polluted models than the canonical ones, could also be explained
by stellar winds unaccounted for by the model atmospheres.
As discussed in \citetalias{Moni07a}, this hypothesis better
explains the too low masses observed for many targets in this
temperature range (see \S \ref{capresultsmass}), because evolutionary
effects or enhanced helium abundance would cause both lower gravity
and higher luminosity, and the calculated mass would agree with
expectations.

\begin{figure}
\begin{center}
\resizebox{\hsize}{!}{\includegraphics{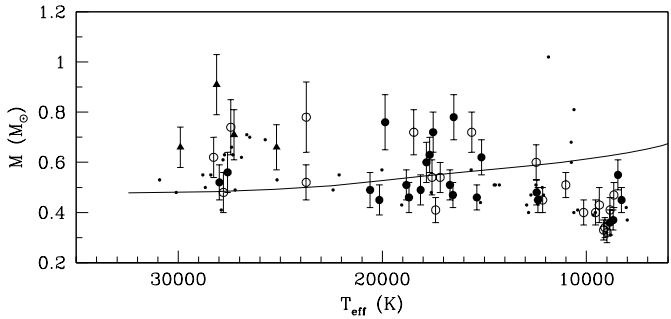}}
\caption{Calculated masses of program stars as function of effective
temperature. Symbols are as in Fig.~\ref{tgplot}. {M\,80 stars which
show anomalous masses are plotted as full triangles.} Errorbars on
temperatures are omitted for clarity. Also the theoretical HB
\citep{Moehler02} is indicated.}
\label{massplot}
\end{center}
\end{figure}

\begin{figure}
\begin{center}
\resizebox{\hsize}{!}{\includegraphics{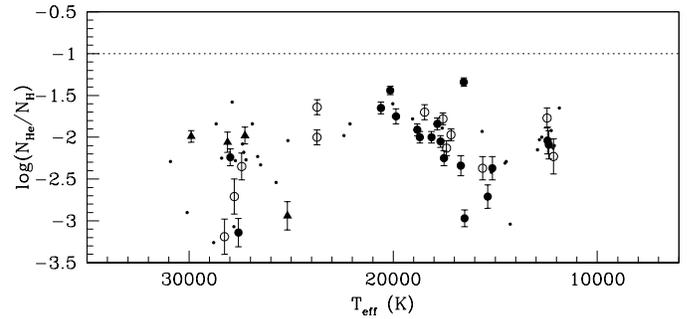}}
\caption{Measured surface helium abundance for program stars. Symbols
are as in Fig.~\ref{tgplot}. Errors in temperature are omitted for
clarity. The dotted line indicates the solar value.}
\label{heplot}
\end{center}
\end{figure}

\subsection{Masses}
\label{capresultsmass}

Our results on masses resemble what found on \object{NGC\,6752}
and discussed in \citetalias{Moni07a}, and are plotted in
Fig.~\ref{massplot}. In brief, masses are systematically lower
than theoretical prediction for stars cooler than 10\,000~K and
in the range T$_\mathrm{eff}$=12\,000-15\,000~K, while they agree
with models for T$_\mathrm{eff}$=15\,000-23\,000~K. Beyond
T$_\mathrm{eff}$=23\,000~K some stars are "normal", while others
show an anomalously high mass.

For temperatures below 10\,000~K masses are too low in both
clusters, as found in \object{NGC\,6752}, despite the fact that the gravities in
Fig.~\ref{tgplot} agree better with the canonical tracks. We conclude
that the problem of low masses in this temperature range cannot be
fully explained only by erroneously low gravities. Between 15\,000
and 23\,000~K, masses fairly scatter around the theoretical track.
As already discussed in \citetalias{Moni07a}, this favors
evolutionary effects or enhanced helium as explanation of the
low gravities measured for many stars among 15\,000 and
17\,000~K, and it argues against the hypothesis of an enhanced
stellar wind (unaccounted for in the models used for parameter
measurements), that would cause underestimated masses.

\begin{figure*}
\begin{center}
\includegraphics[width=16cm]{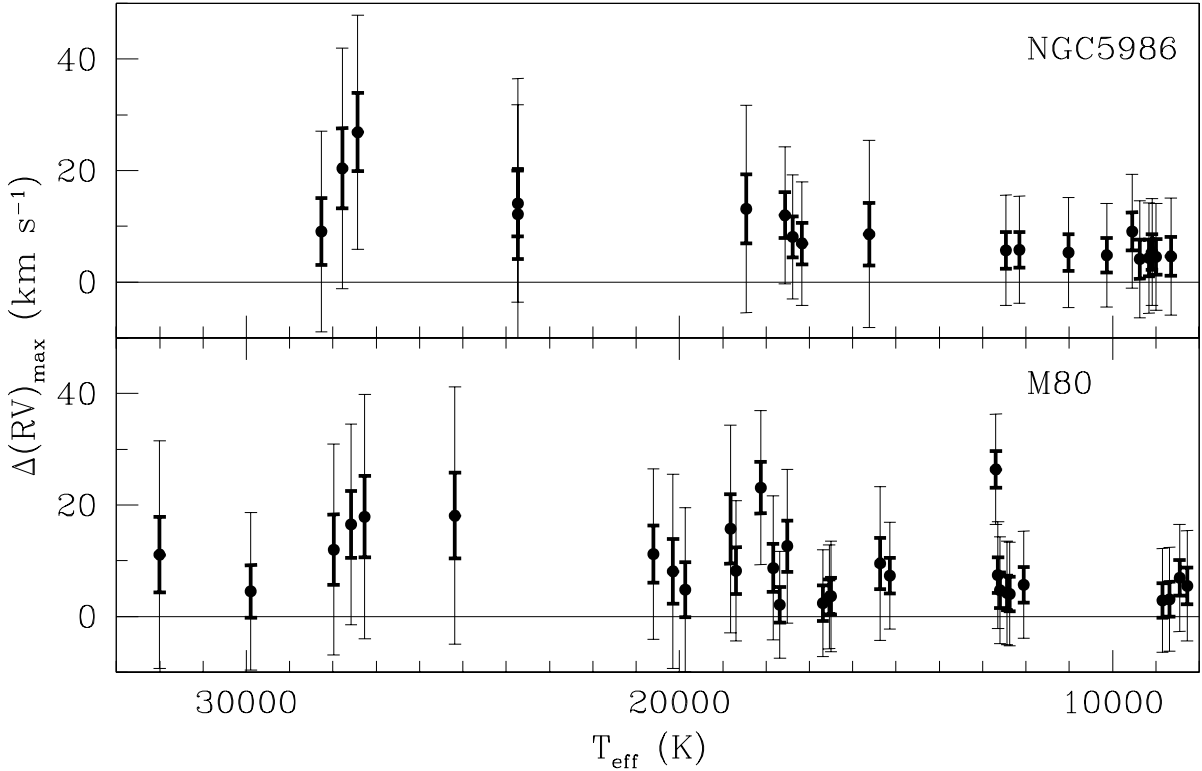}
\caption{Maximum radial velocity variation observed for program
stars. Thin errorbar indicates the 3$\sigma$ interval.}
\label{drvplot}
\end{center}
\end{figure*}

Above 23\,000~K we cannot draw strong conclusions as done in
\citetalias{Moni07a}, because of the small number of stars and
higher errors. In particular, in \object{NGC\,5986} we observed only
five targets, and they are too scattered in the color-magnitude diagram
to try to divide them in different families. On the other hand, in
\object{M\,80} we can confirm what we found in \object{NGC\,6752}.
In fact, we overestimate masses with respect to theoretical predictions
for four out of six stars, and all of them are systematically redder and
slightly fainter than the two for which masses are "normal". These stars
are indicated as full triangles in all figures. Hence, the
strange dichotomy extensively discussed in \citetalias{Moni07a} is
present also in this cluster. We remind the reader that this behavior
of masses is not a consequence of photometric data, because stars with
higher derived masses are fainter, while a lower luminosity alone
would imply a lower calculated mass. Hence, these stars appear both
photometrically {\it and} spectroscopically distinct with respect to
stars showing normal masses. The explanation for this behavior is still
obscure. Higher masses are expected for EHB stars formed through a
merging event of two He white dwarfs \citep{Han02}, but predictions
are still much lower than measured values. We believe the too high masses
are not physical, but just an effect of a mismatch between real stars
and adopted model atmospheres. The same model atmospheres give good mass estimates
for some stars and bad for others, suggesting that the actual stellar atmospheres
are intrinsically different between the two groups of EHB stars.

\subsection{Helium abundance}
\label{capresultshe}

The helium abundance was kept fixed to solar value for stars cooler
than about 11\,000~K, which showed no evidence of diffusion (see \S
\ref{capmeasureparam}). Results for hotter stars are
shown in Fig.~\ref{heplot}. The plot reveals a clear trend with
effective temperature followed by all stars in the three clusters,
although it passed unnoticed by \citetalias{Moni07a}
because masked by the lack of stars between 15\,000 and 20\,000~K
(see their Figure 8).
The helium abundance turns to sub-solar values at 12\,000~K, possibly
reaching a minimum at 15\,000-16\,000~K, then rises again steeply 
and continuously up to about 22\,000~K. Finally helium abundance
decreases again, but this transition region is undersampled. For
stars hotter than 25\,000~K the helium abundances scatter between
$-3 \leq \log{\frac{\mathrm{N_{He}}}{N_H}} \leq -2$.
They possibly follow a double-peaked distribution rather
than a wide single one, but this can be only guessed from the data.
The same general trend can be seen also in
\citet[][their Fig.~5]{Moehler03} and, concerning the first decrease
between 12\,000 and 15\,000~K, in \citet{Behr03} and \citet{Fabbian05},
while it was probably hidden by larger errors in \citet{Moehler00}.
Helium depletion caused by diffusion is usually coupled
with metal overabundances and, interestingly enough, even the iron abundances
measured by \citet[][their Fig.~4]{Pace06} in \object{NGC\,2808} follow
the same trend observed here, reaching a maximum at about 15\,000~K and
then decreasing at higher temperatures.

The observed behavior of helium surface abundances with temperature
does not come unexpected. On the contrary, it confirms theoretical
expectations based on our current understanding of diffusion
processes in the atmospheres of HB stars. Early results of
\citet{Glaspey89} and later detailed studies by many authors (see \S
\ref{capintro}) showed that surface abundances abruptly change at
T$_\mathrm{eff}\approx$11\,000~K. 
The most common interpretation of the observed phenomena relies on
the disappearance of the \ion{He}{i} convection zone at this effective
temperature, that would be responsible for the gravitational settling
of helium, with its consequent depletion in a thin radiative layer between
the surface and the region of \ion{He}{ii} ionization. Within this picture,
this layer becomes thinner for increasing temperatures, because the second He
ionization region moves outward \citep{Sweigart00}, and moreover mass-loss
\citep[which contrasts diffusion,][]{Michaud86} increases
\citep{Vink02}. As a net result, the efficiency of diffusion would be
expected to decrease with increasing temperature, as also suggested
by the photometric properties of stars which, after the Grundahl jump,
progressively reduce the discrepancies with standard models.
This is exactly what we are observing between 15\,000
and 23\,000~K. 
Unfortunately, this scenario has recently been ruled out by the latest
theoretical models, which show that diffusion affects layers much
deeper than the
depth of the \ion{He}{i} convection zone \citep[][see their Figure~5]{Michaud08}.
Actually, the models that best reproduce the observed trend of surface metal
abundances assume that the outer regions down to
$\frac{\mathrm{M}}{\mathrm{M}_\odot}$=10$^{-7}$ are completely mixed by turbulence,
well below the \ion{He}{i} ionization zone (Sweigart, priv. comm.).
In other words, turbulence from this convective layer is not able to inhibit levitation
in Michaud's models. \citet{Michaud08} tentatively attribute the onset of
levitation at $\approx$11\,000~K to an abrupt change in stellar rotational velocity
as observationally confirmed \citep[e.g.][]{RecioBlanco02,Behr03}, but the issue is
currently under debate.

\citet{Momany02} proposed a new onset of diffusion
at about 23\,000~K, causing the photometrical anomaly known as
"Momany Jump". Unfortunately the behavior of helium abundance among
hotter stars is hard to decipher. Both our present results and previous ones
on \object{NGC\,6752} 

\begin{table*}[h!]
\begin{center}
\caption{Results for program stars. Columns 1: ID as in Table~\ref{tabdata}.
Columns 2-7: absolute radial velocity, fundamental parameters (effective temperature,
surface gravity, helium abundance and mass) and maximum radial velocity variation.}
\label{tabres}
\small
\begin{tabular}{ c r r r r r r}
\hline \hline
ID&$V_\mathrm{rad}$&$\mathrm{T_{eff}}$&$\log{g}$&log($\frac{N_{He}}{N_H}$)&M&$\Delta$(RV)$_\mathrm{max}$\\
&km s$^{-1}$&K&&&M$_{\odot}$&km s$^{-1}$\\
\hline
\multicolumn{7}{c}{NGC\,5986} \\
\hline
17512&75$\pm$6&17400$\pm$400&4.38$\pm$0.07&$-$2.13$\pm$0.09&0.41$\pm$0.05&$8.1\pm3.7$\\
17604&--&8850$\pm$150&3.25$\pm$0.07&$-$1.00$\pm$0.00&0.41$\pm$0.05&--\\
17691&94$\pm$6&23700$\pm$1200&5.52$\pm$0.12&$-$1.64$\pm$0.09&0.78$\pm$0.14&$12.2\pm8.1$\\
18077&81$\pm$6&9350$\pm$130&3.38$\pm$0.14&$-$1.00$\pm$0.00&0.43$\pm$0.07&$4.1\pm3.5$\\
18240&76$\pm$6&8650$\pm$90&3.26$\pm$0.05&$-$1.00$\pm$0.00&0.47$\pm$0.05&$4.6\pm3.5$\\
2103&87$\pm$5&12150$\pm$150&3.84$\pm$0.05&$-$2.23$\pm$0.21&0.45$\pm$0.05&$5.8\pm3.2$\\
3192&80$\pm$5&17600$\pm$400&4.41$\pm$0.07&$-$1.78$\pm$0.07&0.54$\pm$0.07&$12.0\pm4.1$\\
3560&92$\pm$7&17200$\pm$300&4.53$\pm$0.05&$-$1.97$\pm$0.07&0.54$\pm$0.06&$6.9\pm3.7$\\
4175&93$\pm$11&27400$\pm$800&5.44$\pm$0.10&$-$2.35$\pm$0.16&0.74$\pm$0.11&$26.9\pm7.0$\\
4930&98$\pm$5&10100$\pm$160&3.48$\pm$0.07&$-$1.00$\pm$0.00&0.40$\pm$0.05&$4.8\pm3.1$\\
5558&84$\pm$6&12500$\pm$150&3.90$\pm$0.05&$-$1.77$\pm$0.12&0.60$\pm$0.07&$5.7\pm3.3$\\
6102&93$\pm$9&15600$\pm$300&4.30$\pm$0.05&$-$2.37$\pm$0.14&0.72$\pm$0.08&$8.6\pm5.6$\\
7008&81$\pm$5&9000$\pm$160&3.15$\pm$0.08&$-$1.00$\pm$0.00&0.32$\pm$0.04&$4.5\pm3.2$\\
7430&90$\pm$7&9550$\pm$140&3.37$\pm$0.07&$-$1.00$\pm$0.00&0.40$\pm$0.05&$9.1\pm3.4$\\
8131&89$\pm$7&28300$\pm$700&5.37$\pm$0.08&$-$3.19$\pm$0.21&0.62$\pm$0.08&$9.1\pm6.0$\\
9049&92$\pm$6&18500$\pm$500&4.67$\pm$0.07&$-$1.70$\pm$0.09&0.72$\pm$0.09&$13.1\pm6.2$\\
9250&91$\pm$6&9150$\pm$140&3.21$\pm$0.08&$-$1.00$\pm$0.00&0.33$\pm$0.04&$4.3\pm3.3$\\
10390&89$\pm$7&9100$\pm$130&3.12$\pm$0.07&$-$1.00$\pm$0.00&0.34$\pm$0.04&$5.4\pm3.2$\\
11215&91$\pm$7&11000$\pm$110&3.67$\pm$0.03&$-$1.00$\pm$0.00&0.51$\pm$0.05&$5.3\pm3.3$\\
11571&102$\pm$10&27800$\pm$900&5.53$\pm$0.12&$-$2.71$\pm$0.21&0.48$\pm$0.08&$20.4\pm7.2$\\
12099&103$\pm$11&23700$\pm$800&5.14$\pm$0.08&$-$2.00$\pm$0.09&0.52$\pm$0.07&$14.1\pm5.9$\\
\hline
\multicolumn{7}{c}{M\,80} \\
\hline
14327&15$\pm$7&20100$\pm$700&5.00$\pm$0.08&$-$1.44$\pm$0.05&0.45$\pm$0.06&$8.1\pm5.8$\\
14786&6$\pm$4&8280$\pm$50&3.12$\pm$0.02&$-$1.00$\pm$0.00&0.45$\pm$0.05&$5.5\pm3.3$\\
14985&6$\pm$9&17500$\pm$400&4.78$\pm$0.05&$-$2.25$\pm$0.09&0.72$\pm$0.08&$12.6\pm4.6$\\
15200&12$\pm$6&17800$\pm$500&4.67$\pm$0.07&$-$1.84$\pm$0.07&0.60$\pm$0.08&$8.7\pm4.3$\\
16389&3$\pm$9&18100$\pm$400&4.75$\pm$0.05&$-$2.00$\pm$0.07&0.49$\pm$0.06&$23.1\pm4.6$\\
16163&5$\pm$5&15400$\pm$200&4.25$\pm$0.05&$-$2.71$\pm$0.14&0.46$\pm$0.05&$9.5\pm4.6$\\
17173&10$\pm$13&12700$\pm$300&4.23$\pm$0.12&$-$1.00$\pm$0.00&2.40$\pm$0.39&$26.4\pm3.3$\\
17114&8$\pm$4&16500$\pm$300&4.45$\pm$0.05&$-$2.97$\pm$0.10&0.78$\pm$0.09&$3.6\pm3.3$\\
16707&4$\pm$7&--&--&--&--&$11.1\pm6.8$\\
17737&1$\pm$5&28100$\pm$700&5.69$\pm$0.08&$-$2.06$\pm$0.12&0.91$\pm$0.12&$11.8\pm6.1$\\
18516&10$\pm$5&18700$\pm$500&4.78$\pm$0.07&$-$2.00$\pm$0.07&0.46$\pm$0.06&$8.2\pm4.2$\\
18110&10$\pm$4&12400$\pm$150&3.84$\pm$0.03&$-$2.04$\pm$0.16&0.48$\pm$0.05&$4.2\pm3.1$\\
18992&7$\pm$4&8850$\pm$180&3.16$\pm$0.08&$-$1.00$\pm$0.00&0.36$\pm$0.05&$2.9\pm3.1$\\
19040&15$\pm$5&18800$\pm$400&4.81$\pm$0.05&$-$1.91$\pm$0.07&0.51$\pm$0.06&$15.7\pm6.2$\\
18391&12$\pm$3&12100$\pm$200&4.37$\pm$0.07&$-$1.00$\pm$0.00&5.13$\pm$0.64&$5.7\pm3.2$\\
19191&13$\pm$9&20600$\pm$600&5.08$\pm$0.07&$-$1.65$\pm$0.07&0.49$\pm$0.07&$11.2\pm5.1$\\
19127&13$\pm$11&25200$\pm$700&5.51$\pm$0.08&$-$2.94$\pm$0.17&0.66$\pm$0.09&$18.1\pm7.7$\\
12304&7$\pm$6&27300$\pm$600&5.65$\pm$0.08&$-$1.98$\pm$0.10&0.71$\pm$0.10&$17.9\pm7.3$\\
14201&16$\pm$5&16700$\pm$400&4.43$\pm$0.05&$-$2.34$\pm$0.12&0.51$\pm$0.06&$2.4\pm3.2$\\
13787&1$\pm$4&17700$\pm$300&4.72$\pm$0.05&$-$2.05$\pm$0.07&0.63$\pm$0.07&$2.1\pm3.2$\\
12663&19$\pm$5&27600$\pm$700&5.38$\pm$0.08&$-$3.14$\pm$0.17&0.56$\pm$0.08&$16.5\pm6.0$\\
12767&20$\pm$9&29900$\pm$500&5.73$\pm$0.05&$-$1.99$\pm$0.07&0.66$\pm$0.08&$4.5\pm4.7$\\
13839&$-$4$\pm$11&8450$\pm$40&3.26$\pm$0.02&$-$1.00$\pm$0.00&0.55$\pm$0.06&$6.9\pm3.2$\\
14022&$-$11$\pm$9&19900$\pm$800&5.15$\pm$0.08&$-$1.75$\pm$0.09&0.76$\pm$0.11&$4.8\pm4.9$\\
14387&1$\pm$8&15100$\pm$300&4.36$\pm$0.05&$-$2.37$\pm$0.14&0.62$\pm$0.07&$7.3\pm3.2$\\
12526&6$\pm$15&28000$\pm$700&5.51$\pm$0.08&$-$2.24$\pm$0.10&0.52$\pm$0.07&$12.0\pm6.3$\\
13179&15$\pm$7&16500$\pm$300&4.17$\pm$0.05&$-$1.34$\pm$0.05&0.47$\pm$0.05&$3.5\pm3.1$\\
15682&14$\pm$10&12400$\pm$200&3.78$\pm$0.05&$-$2.09$\pm$0.17&0.45$\pm$0.05&$4.0\pm3.1$\\
14327&15$\pm$7&20500$\pm$500&5.06$\pm$0.07&$-$1.42$\pm$0.05&0.49$\pm$0.06&$5.8\pm4.5$\\
15183&0$\pm$15&12600$\pm$400&4.19$\pm$0.12&$-$1.00$\pm$0.00&3.09$\pm$0.06&$7.4\pm3.2$\\
14813&15$\pm$9&8700$\pm$100&3.16$\pm$0.05&$-$1.00$\pm$0.00&0.37$\pm$0.04&$3.1\pm3.1$\\
15470&8$\pm$6&12600$\pm$300&4.23$\pm$0.12&$-$1.00$\pm$0.00&2.95$\pm$0.48&$4.7\pm3.2$\\
\hline
\end{tabular}
\end{center}
\end{table*}
\clearpage

{\noindent
\citep{Moni07a,Moehler00} show that
helium is depleted between a factor of 10 and 100 for these stars,
confirming that diffusion is active at these temperatures, but
with a scatter that is much larger than observational errors.
Stars with the same temperature and different helium depletion seem
to co-exist. Even evolutionary effects cannot be ruled out, as helium
surface abundance could be a time-dependent result of competing
processes \citep{Michaud83}.
As already noted in \citetalias{Moni07a}, the helium abundances of EHB stars
are not related with anomalous masses  discussed in \ref{capresultsmass}.
For example, stars \#12663 and \#19127 are strongly depleted in
helium (log(N(He)/N(H))$\approx-3$), but the mass is "normal" for the first
and too high for the second one.
}

The comparison of our results with Figure~1 of \citet{OToole08} can
be very instructive. In that plot the author summarizes current knowledge
about the trend of helium abundance with temperature for field sdB stars,
gathering the data from many extensive surveys
\citep{Edelmann03,Lisker05,Stroer05,Hirsch08}. The aim of \citet{OToole08}
is to analyze the two families of field EHB stars discovered by
\citet{Edelmann03}, one being a factor of ten more depleted in helium
than the other. Field stars cooler than
20\,000~K are undersampled, but they clearly do not follow the trend
observed by us in three globular clusters. On the contrary, the helium abundances
in the figure of O'Toole seem to decrease monotonically from 12\,000~K to at
least 23\,000~K. We have no explanation for this difference, but we
think it deserves further investigation.

At a first glance, results for globular cluster EHB stars agree
fairly well with those for field EHB stars,
with a possible 
small decrease of helium abundance between 20\,000 and
23\,000~K, and hotter stars being scattered between
$\log(\frac{\mathrm{N_{He}}}{N_H})=-$1.8 and $-$3.2. Hence, quite
surprisingly, we would not be observing any evidence of a second, more
depleted family of stars. These are only about 15\% of field EHBs in the
studied temperature range, and it might be possible to explain their
absence in our sample by their relative scarcity alone.
Nevertheless, we would statistically expect about 4 such
stars in our sample. If these He-poorer objects were stars with no core
He-burning, as proposed by \citet{OToole08}, there would be no clear reason
for their absence in globular clusters. However, at a further inspection an
alternative
interpretation is possible. In fact, comparing the diagrams one could suspect
that an offset is present for EHB stars, with our abundances being
higher by about
0.5 dex.
For example, our stars with T$_\mathrm{eff}$=20\,000~K cluster at
about $\log{\frac{\mathrm{N_{He}}}{N_H}}=-$1.5, while in Figure~1 of
\citet{OToole08} they are at about $-$2. Lowering our abundances by
0.5 dex in Fig.~\ref{heplot}, the bulk of EHB stars would lie between
$\log{\frac{\mathrm{N_{He}}}{N_H}}=-$2 and $-$3, while
five stars would be much more depleted, at
about $\log{\frac{\mathrm{N_{He}}}{N_H}}=-$3.7.
In this case, the helium abundance of both the main population and He-depleted
stars, and even the number ratio of the two families of EHB stars,
would excellently agree with results among sdB stars. However,
arguing for the presence of this offset is quite speculative, because
stars cooler than 20\,000~K show no evidence of it. The procedure adopted
here to measure helium abundance is very standard for sdB studies, and we share
the fitting routine with almost all field surveys. Even the model atmospheres
are often the same \citep[e.g.][]{Edelmann06}, and no offset has ever
been observed. To verify an {\it intrisic} higher helium abundance
for our cluster EHB stars would require a more extensive sample.
As a conclusion, the presence of a helium-poorer EHB population in globular clusters,
analogous to what is observed among field stars, must currently remain an open
issue.

Comparing our results with theoretical expectations is difficult,
because the effects of diffusion processes still lack a full comprehension.
Recent calculations of \citet{Michaud08} well reproduce observed surface
abundances, but their trend with effective temperature (that is our
observational result) is still unexplored.
Moreover, a detailed model prediction cannot neglect the counter-acting
effects of stellar wind, which are still poorly known. In fact,
they are suspected to play an important role, because helium abundances
among sdB stars are much too high to be accounted for by diffusion models
\citep{Michaud89}, and weak stellar winds can explain the discrepancy
\citep{Fontaine97,Unglaub98}.
The combination of diffusion and stellar wind could also produce time-dependent
surface abundances, although on a relatively short time scale, that could affect
the results by introducing an intrinsic star-to-star scatter.
A detailed comparison of our results with field sdB stars would also require a
precise knowledge of the effects of metallicity. In fact, field sdBs should on
average be younger and more metal-rich than our sample.
\citet{Unglaub08} showed that strong coupled stellar winds, involving H and He
in addition to accelerated metals, are prevented for metal-poor EHB stars. The
effects of this result on the helium surface abundance is not stated by the author,
but it could be an important clue to interpret the differences between our results
and the plot of \citet{OToole08}.


\section{Radial velocity variations}
\label{capresultsrvar}

The arc lamp spectra for star \#17604 in \object{NGC\,5986}
were damaged by lines of hot pixels, resulting in a wavelength calibration
without the precision required by
our aims. The target is very cool, adding little information to our
search for binaries focused on EHB stars, and we considered
safer to exclude it from analysis.

Our radial velocity results are summarized in Fig.~\ref{drvplot},
where we plot the maximum variation for each star. Variations are
taken always positive, the sign just being a consequence of the
arbitrary definition of "template" and "object" spectrum. We indicate
the 3$\sigma$ interval with thinner errorbars.

We detect RV variations above 3$\sigma$ for one of the four cool
targets suspected to be foreground main sequence stars in \object{M\,80}
(\S \ref{capresultsparam}), indicating it could be a spectroscopic binary.
We did not find evidence for a companion in our spectra, although at
our low resolution its features could be easily hidden by the lines of
the brighter primary. We found that all weak lines, in particular Fe lines
and the MgIb triplet, follow H$_{\beta}$ in its RV variations.
Therefore, the companion could be a compact object or a low-mass main sequence star.
No other cool stars shows any variability within our errors. Hence,
no close binary is detected among them, in agreement with recent
results from a large sample of cool HB stars 
in \object{NGC\,6752} \citep[104 HB stars
with T$_\mathrm{eff}\leq$20\,000~K,][]{Moni08}.

Star \#14327 shows no noticeable RV variation in both field M\,80a and M\,80b.
We also checked that no variation occurs between its spectra in the two
fields. Nevertheless, it is the only star observed seven times, and in
order to avoid in the statistical analysis one star with
different temporal sampling with respect to all the others, we will
simply exclude results from slit 12b, thus considering this star as
a normal target observed in field M\,80a only. The choice of the slit
to exclude is dictated by the better temporal sampling of M\,80a field.

Among the EHB stars we detect one interesting close binary candidate for
each cluster, namely star \#4175 in \object{NGC\,5986} (maximum
variation 26.9$\pm$7.0 km s$^{-1}$) and \#16389 in \object{M\,80}
(maximum variation 23.1$\pm$4.6 km s$^{-1}$). Variations for these
targets are small compared to those observed on typical sdB close
binaries \citep[see for example][]{Maxted01,MoralesRueda03}, but
so are the errors. In particular, CC errors for these measurements are
very small compared with stars with similar S/N. In our experience, this
could be an indirect indication that the variation is real and not
due to noise-induced distortion of the CC function.
Some stars show variations at the edge of the
$3\sigma$ interval, although not formally outside it, and we want to verify
if other candidates could hide among them.
We performed statistical calculations on all EHB targets to translate these
considerations into numbers. Results are summarized in
Table~\ref{stattab}. The star IDs and the maximum observed RV variations are
indicated in column~1 and 2, respectively. In column~3 the maximum RV
variation is given in units of $\sigma$.
To estimate the significance of variations near the 3$\sigma$ threshold,
in column 4 we calculated the probability that a variation greater or
equal to $\frac{\Delta(\mathrm{RV})_\mathrm{max}}{\sigma}$
occurs among the measurements, assuming a normal distribution of errors.
From basic statistical relations, we have:
\begin{equation}
p\big{(}\frac{\Delta(\mathrm{RV})_\mathrm{max}}{\sigma}\big{)}=
1-\big{[}\mathrm{erf}\big{(}\frac{\Delta(\mathrm{RV})_\mathrm{max}}{\sqrt{2}\sigma}
\big{)}\big{]}^{n},
\label{eq2}
\end{equation}
where $n$ is the number of independent measurements (10 in NGC\,5986,
6 in M\,80a, and 3 in M\,80b), and
\begin{equation}
\mathrm{erf}(x)=\frac{2}{\sqrt{\pi}}\int_{0}^{x}e^{-t^{2}}dt.
\label{eq_erf}
\end{equation}
Finally, we calculated the fraction
of typical sdB close binaries with random phase and inclination angle, a
0.5 M$_\odot$ companion, and period P$\leq$10 days, that in our survey
would show a RV variation {\it not} greater than the maximum observed.
This is an estimate of how likely the small observed variations can
really indicate a close binary. Results are tabulated in column 5.
We included in the analysis one cool
target (\#14985 in \object{M\,80}) with RV variations near
3$\sigma$, marked with an asterisk in Table~\ref{stattab}. 

\begin{table}[t]
\begin{center}
\caption{Statistical analysis of RV variations for our target EHB stars.}
\label{stattab}
\begin{tabular}{c| c c c c}
\hline
ID & $\Delta$(RV)$_\mathrm{max}$ &
$\frac{\Delta(\mathrm{RV})_\mathrm{max}}{\sigma}$
& p($\frac{\Delta(\mathrm{RV})_\mathrm{max}}{\sigma}$)
& p($\leq\Delta$(RV)$_\mathrm{max}$) \\
 & (km s$^{-1}$) & & (\%) & (\%) \\
\hline
\multicolumn{5}{c}{NGC\,5986}\\
\hline
17691 & 12.2 & 1.51 & 73.6 & 11.9 \\
4175 & 26.9 & 3.84 & 0.15 & 28.3 \\
8131 & 9.1 & 1.52 & 73.6 & 8.3 \\
11571 & 20.4 & 2.83 & 4.0 & 21.2 \\
12099 & 14.1 & 2.39 & 13.4 & 14.1 \\
\hline
\multicolumn{5}{c}{M\,80}\\
\hline
14327 & 8.1 & 1.40 & 60.4 & 11.9 \\
14985* & 12.6 & 2.74 & 3.0 & 19.5 \\
16389 & 23.1 & 5.02 & 2.4$\cdot 10^{-6}$ & 35.1 \\
16707 & 11.1 & 1.63 & 44.6 & 35.1 \\
17737 & 11.8 & 1.93 & 26.7 & 18.2 \\
19191 & 11.2 & 2.20 & 14.2 & 17.2 \\
19127 & 18.1 & 2.35 & 10.0 & 28.0 \\
12304 & 17.9 & 2.45 & 3.5 & 51.0 \\
12663 & 16.5 & 2.75 & 1.5 & 48.3 \\
12767 & 4.5 & 0.96 & 65.2 & 17.5 \\
14022 & 4.8 & 0.98 & 65.2 & 18.5 \\
12526 & 12.0 & 1.90 & 14.4 & 38.6 \\
\hline
\end{tabular}
\end{center}
\end{table}

The numbers in Table~\ref{stattab} help to clarify the situation.
First, the two stars that show variations greater than 3$\sigma$ are
confirmed to be good candidates, in view of the negligible
probability of the observed datum being
a random variation. Their variations are indeed small,
but not enough to rule out binarity: about two out
of three typical sdB binaries would show higher variations in our
observations but, for example, this fraction would decrease if we assume
a lower mass companion.

For all other EHB targets, except for field M\,80b because of poor temporal
sampling, the variations
are unlikely to be caused by binarity because of low probabilities
in column 5, although it cannot be excluded on the basis of this datum
alone. The numbers in column 4 are more
conclusive, and they do not allow us to claim the detection of any other
binary candidate. In some cases the probability of the datum
is low, but not negligible. One doubtful exception could be star \#12663
in \object{M\,80}, but the probability of its observed RV variation
being random is still at least one order of magnitude higher than for the two
proposed candidates. Moreover, it is important to notice that
this probability is extremely sensitive to the adopted error value. We
calculated that probabilities lower than 4\% in column 4 would double
if errors were 10\% larger. Despite our efforts, we feel that
our error estimates are unlikely to be more precise than this value.
Therefore these low probabilities should be considered accurate only
to within a factor of two.

In summary, the numbers support the hypothesis of binarity for only 
one star per cluster, the only ones which show variations greater than 3$\sigma$.
This indicates that requiring a variation greater than 3$\sigma$ for the
detection of a binary system is a good choice, and in the statistical
analysis we will assume this value as detection threshold.

The only other EHB close binary discovered to date in a globular cluster \citep{Moni08}
shows a MgIb triplet anomalously strong for stars
at its temperature \citep{Moni07a}, indicative of a cool main sequence companion. We
inspected the sum of all the V1400 spectra,
but we never found evidence for this feature in any EHB target. Therefore,
the companions of our binary candidates are most likely compact objects
(as white dwarfs), or very low-mass main sequence stars. In fact, we have no indication
about the period, and
very low-mass companions in close nearly edge-on orbits can cause large
RV variations \citep[see for example the NY Vir system,][]{Vuckovic07}.
Thus, a low-mass main sequence companion is plausible.


\section{EHB close binary fraction}
\label{capbinaries}

\begin{figure}
\begin{center}
\resizebox{\hsize}{!}{\includegraphics{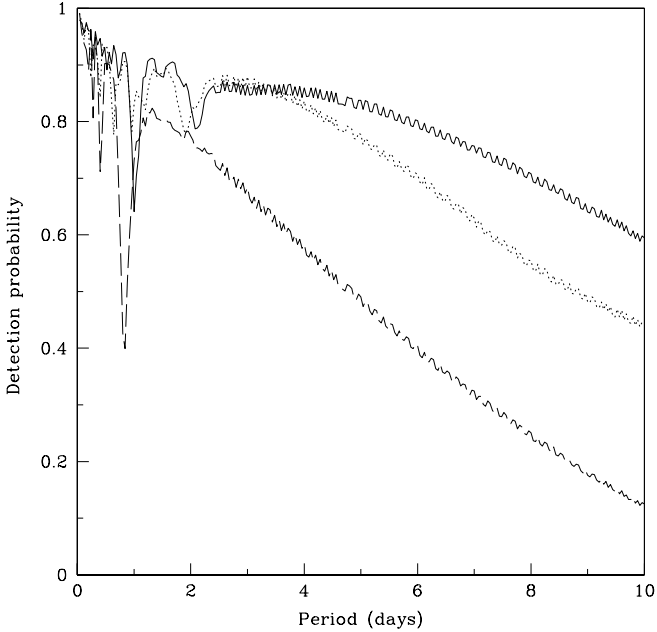}}
\caption{Probability of binary detection in our survey, as a function
of binary period. {\it solid line}: NGC\,5986.
{\it dotted line}: field M\,80a. {\it dashed line}: field M\,80b.
The calculation assumes a companion of 0.5 M$_\odot$ in
circular orbit.}
\label{probplot}
\end{center}
\end{figure}

In the statistical analysis of our results we will assume the
detection of two EHB close binaries, one per cluster, as discussed
in \S \ref{capresultsrvar}. The candidate in \object{M\,80} is
slightly cooler than the typical temperature boundary
for EHB stars (T$_\mathrm{eff}\geq$20\,000~K).
The distinction is more than just a conventional definition,
because it approximately sets the transition between classical HB stars
with envelopes sufficiently massive to sustain the hydrogen-burning
shell, and EHB stars, which do not have this second energy 
source. The post-HB evolution
is also very different, as depicted in \S\ref{capintro}. It can also
be noted from Fig.~\ref{cmd} that at approximately 20\,000~K there
is an underpopulated region in the HB of both clusters, and the binary
in \object{M\,80} is brighter than this gap (at V$\approx$18.7),
indicating that it is indeed cooler than typical EHB stars.
Despite these arguments, we consider
that excluding the candidate from statistical analysis on the basis
of its temperature alone would be quite artificial. In fact, field
surveys usually focus on hotter stars, excluding targets with lower
temperatures \citep[see for example][]{Maxted01}, but they are
sometimes included \citep{Ulla98}, and binary systems are detected
among them \citep{Aznar01}. Population synthesis models also indicate
that sdBs as cool as 15\,000-16\,000~K can be formed by interactions
within binary systems \citep{Han03}. Moreover, relatively massive
(0.75 M$_\odot$) EHB stars, or with a relatively massive envelope
($\geq$0.01 M$_\odot$), move to temperatures lower than 20\,000~K
during the first stages of post-EHB evolution \citep{Han02}. Hence,
the candidate could even be an evolving object.
Its exclusion would strengthen the results in favor of a lack of
close systems.

In the previous section we found that some hot stars show  
marginally significant RV variations. We emphasize that 
the following analysis does not depend on whether we consider them
as candidates or not. In fact, once the detection threshold is fixed,
the only input is the number of stars with variations above and below
it, and the routines automatically consider undetected systems
(because of unfavorable temporal sampling or low inclination angles) as
part of the calculations.

The close binary detection probability of our survey was calculated as
in \citetalias{Moni06a}. In brief, 2500 typical sdB binaries
\citep[with a 0.5 M$_{\odot}$ companion in circular orbit, as
assumed for example by][]{Maxted01,MoralesRueda06} were simulated for
each value of period P, evenly distributed in the phase-$\sin{i}$ space
(where i is the inclination of orbit with respect to the line of sight).
Then we calculated the fraction of these synthetic systems that
would have been detected in our
observations, i.e. showing RV variation greater than the detection
threshold. We fixed the threshold at the 3$\sigma$ value,
because we found (\S \ref{capresultsrvar}) that variations
lower than this value are not sufficiently significant. We adopted
3$\sigma$=20 km s$^{-1}$ for
\object{NGC\,5986} and 18 km s$^{-1}$ for \object{M\,80}, which are
average values for our program stars and well represent the typical
accuracy of measurements. Results are plotted in Fig.~\ref{probplot}.
The detection probability for \object{NGC\,5986} is high, as a consequence
of the good temporal sampling. On the contrary,
the lack of data in the first two nights for \object{M\,80}, due to bad
weather, strongly damaged the efficiency of the survey, mainly in the field
M\,80b. We will limit our analysis to periods P$\leq$5 days on this cluster,
as done by \citet{Moni08} for similar reasons. This limitation is not
too severe, because our main aim is the comparison with results on
field sdB stars. The exact shape of their period distribution is still
unknown, but is suspected to be strongly peaked at P=1day, and binaries
with P$\geq$5 days represent only the tail of the distribution
\citep[see for example Fig.~2 of][]{MoralesRueda06}.

\begin{figure}
\begin{center}
\resizebox{\hsize}{!}{\includegraphics{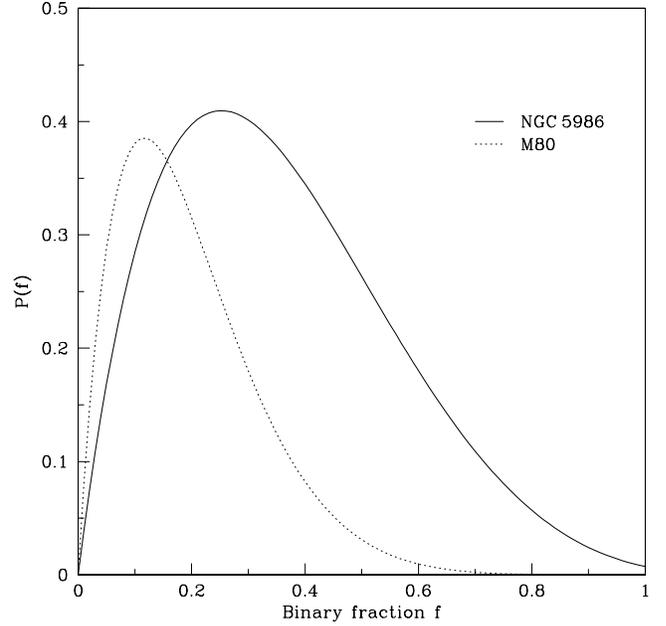}}
\caption{Curves of probability for the close binary fraction $f$,
as calculated from our results. {\it solid line}: NGC\,5986
(binaries with periods P$\leq$10 days). {\it dotted line}: M\,80
(periods P$\leq$5 days).}
\label{fracplot}
\end{center}
\end{figure}

The probability of detecting N$_B$ binaries out of a sample of N targets
is:
\begin{equation}
\mathrm{p} = \frac{\mathrm{N!}}{(\mathrm{N-N_{\mathrm{B}}})!\mathrm{N_{\mathrm B}}!}
(\mathrm{\bar{d}}f)^\mathrm{N_{\mathrm B}}(1-\mathrm{\bar{d}}f)^{\mathrm{N-N_{\mathrm B}}},
\label{eq_prob}
\end{equation}
where $f$ is the binary fraction and $\bar{d}$ is the probability
of detection weighted with the period distribution. For \object{M\,80} we
used the mean of the detection probability of the two fields,
weighted with the number of EHB
targets observed in each. The shape of the period distribution
affects the results only marginally, as already demonstrated by
\citetalias{Moni06a} and \citet{Moni08}. A Gaussian distribution in
$\log{P}$ centered on $\log{P}$=0 days, as proposed by \citet{Maxted01} and
\citet{Napiwotzki04}, does not change p($f$) by more than 0.01-0.03 with
respect to the flat distribution we assumed here.
We used the relation (\ref{eq_prob}) as a function of $f$ to calculate the
probability p($f$), given our observational results of one detection out
of five targets in \object{NGC\,5986} and one out of eleven
in \object{M\,80}. As stated before, we must limit our analysis to periods
P$\leq$5 days on \object{M\,80}. The results of our calculations are plotted in
Fig.~\ref{fracplot}.
The curves are far from being Gaussian-shaped, hence we cannot simply
deduce a best estimate with an associated error. Nevertheless, from their
analysis we can draw important conclusions.

\subsection{NGC\,5986}
\label{cap5986bin}

In \object{NGC\,5986} the most probable value of
$f_\mathrm{P\leq 10days}$ is 25\%, as indicated by the peak of the curve
in Fig.~\ref{fracplot}. This must be considered the best estimate for the
EHB close binary fraction in this cluster. Nevertheless, the small number
of targets implies a very wide curve, so that no value of $f$ can be safely
excluded. A very high binary fraction as proposed by \citet{Maxted01}
for field stars is not probable, but cannot be completely ruled out
(p($f$=0.7)=10.9\%). The probability of our best estimate is
p($f$=0.25)=41\%, which does not noticeably differ from p($f$=0.4).
Hence, a binary fraction as high as the lowest determination among field
sdB stars is not preferred, but it is perfectly reasonable.
There is no improvement in limiting to periods shorter than
5 days, because the width of the probability curve is dictated
by small number statistic and not by the sensitiveness of the survey.
We conclude that our data do not give strong constraints on the EHB 
close binary fraction for this cluster, and they agree both with the
unexpected low estimates found in \object{NGC\,6752} \citep{Moni06a,Moni08},
{\it and} with the lowest determinations for field sdBs
\citep{Napiwotzki04}. They tend to exclude high binary fractions
as proposed by \citet{Maxted01} for field sdBs.
Given the most probable value, the results hint that the real
fraction could be relatively low.

\subsection{M\,80}
\label{cap80bin}

Stronger results can be obtained on \object{M\,80}. The best estimate
is $f_\mathrm{P\leq 5days}$=12\%, very low compared to any determination
among field stars. High values are ruled out (p($f$=0.7)=0.2\%).
Results well agree with the extremely low fraction found by
\citet{Moni08} in \object{NGC\,6752} (p($f$=0.04)=25\%), much more than
the lowest values for field sdBs, which are very improbable for this
cluster (p($f$=0.4)=8.2\%). Within a 90\% confidence level, $f$ is lower
than 38\%.

In summary, although our results are not as strong as previous ones on
\object{NGC\,6752}, we find that {\it also in \object{M\,80} EHB close binary
systems are lacking}, at variance with what is observed
among field sdBs. The binary fraction is not very well constrained,
but values observed among field samples are very unlikely, while
results agree well with the tiny 4\% found in \object{NGC\,6752}
\citep{Moni08}. \object{M\,80} is the second globular cluster for which a lack of EHB
close systems is found, and this indicates that it should not be
a peculiarity of \object{NGC\,6752}. Preliminary results by
\citet{Moni07b} suggest that some clusters could be different, but the
early stage of their analysis and their small number statistics strongly
call for further investigation. In \S \ref{capintro} we discussed the
models for the still unclear sdB star formation mechanisms. In light of
these results, any successful model must take into account the significant
difference between stars inside and outside globular clusters.

\subsection{Comparison with Han (2008) results}
\label{capHanbin}

As mentioned in \S \ref{capintro}, \citet{Han08} confirmed with
theoretical calculations that the binary scenario naturally implies a
$f$-age relation, as proposed by \citet{Moni08}. In brief,
the model assumes that dynamical interactions in binary systems are
responsible for sdB star formation, but the efficiency of the various
channels varies with the age of sdB progenitors,
leading to a decreasing fraction of {\it close} binaries with increasing
mean age of the population.

The general conclusion of our work, that $f$ is low also in a second
globular cluster and possibly in a third one, is a fair confirmation
of Han's model. The fraction predicted in \object{NGC\,6752} by
his preferred set of parameters (2\%)  fairly agrees
with observational results of \citet{Moni08}. Unfortunately, a direct
comparison for the two new globular clusters is complicated by the large
uncertainties in both model predictions and empirical determinations.
Therefore, the analysis cannot rely on exact values,
but on the general behavior of the results.

The best estimates of $f$ in the three clusters so far surveyed
are all smaller than those found for the field sdB stars,  but
too different from each other if compared to Han's prediction.
The lowest observed fraction is for \object{NGC\,6752} and the
highest is for \object{NGC\,5986}. The order is correct, because
\object{NGC\,6752} is the oldest cluster while \object{NGC\,5986}
is the youngest one \citep{DeAngeli05}, but age difference is too
small, about 2 Gyr, compared with the modeled $f$-age relation.

The parameter set preferred by Han creates an $f$-age relation
that fairly agrees with observations of \object{NGC\,6752}, but
its predictions are not easily compatible with our new results:
the model implies a nearly constant $f$=2\% for populations older
than 10 Gyrs. Despite the uncertainties in our study, it is clear
that this low value is improbable in both \object{NGC\,5986}
(6\% probability) and \object{M\,80} (14\%, see Fig.~\ref{fracplot}).
The combined probability, i.e. the probability that $f\approx$2\%
in both cluster is negligible, according to our observations.
This tends to exclude the $f$-age relation derived by this
model.  Other models studied by \citet[][see his Fig.~3]{Han08},
with different sets of input parameters, predict higher close binary
fractions and a steeper relation with age, thus solving the outlined
contradictions, but their high expected $f$ are incompatible with
measurements in \object{NGC\,6752}, the more robust of the
observed results.

From this analysis we conclude that, despite the good general
agreement (close EHB binaries are predicted and observed to be a
minor population), models and observations still lack a good
agreement on the details, although observational constraints
are still not strong enough to be conclusive.
Maybe the discrepancies could be mitigated by a refined set of
model parameters, or some other effect (like dynamical interactions
in dense environments) could be invoked to slightly change
$f$ from cluster to cluster.

It is important to note that the present comparison strongly relies
on our results on \object{M\,80}, which are the most precise, and
the low temperature of the binary candidate in this cluster
leaves space for doubts on the feasibility of the
comparison itself. In fact, \citet{Han08} did not apply any
temperature cut, but his $f$-age relation was obtained taking into
account a GK selection effect\footnote{The GK selection effect is
an observational bias against EHB stars with a companion of G-K
spectral type, because of the composite spectrum, or earlier,
because of spectral dominance of the companion.}, and models
corrected to consider this bias tend to exclude sdBs as cool as our
target \citep{Han03}. Anyway, the system in \object{M\,80} should not
be a EHB+MS wide binary (\S\ref{capresultsrvar}) as the systems
selected against by the GK effect, so the issue remains uncertain.
\citet{Han08} just states that, in absence of the GK effect,
his predicted $f$ should be smaller at any age.

We conclude that the confirmation and
refinement of \citet{Han08} model require more precise empirical
$f$ measurements in these and other globular clusters, that could even help
constraining the model parameters, in particular the
physically-important and poorly-known common envelope efficiency
$\alpha_{CE}$ \citep[see][for a discussion]{Han08}.


\section{Summary}
\label{capsummary}

We analyzed radial velocity variations for 51 hot HB/EHB stars in two Galactic
globular clusters, in search for signatures of
close binary systems. We also studied low-resolution spectra of program
stars measuring temperature, surface gravity, helium abundance and mass.
Our main results can be summarized as follows:

\begin{itemize}
\item In \object{M\,80} we confirm the anomalous behavior of spectroscopic
masses found in \object{NGC\,6752} \citep{Moni07a} for stars hotter than
23\,000~K, although this result is less evident due to a smaller sample.
Stars being fainter and/or redder show too high masses with respect to
theoretical expectations, whereas their lower luminosities would suggest
lower masses. In \object{NGC\,5986} the small
number of EHB target observed prevents such an analysis.

\item For the first time we observe a clear trend of helium abundance
with temperature along the entire blue HB. These results confirm
that helium depletion due to atmospheric diffusion reaches a maximum
at about 15\,000~K, then the helium abundance rises again with increasing
temperature. This behavior agrees qualitatively with expectations for
the effects of diffusion at different temperatures along the HB.
Somewhere at about 23\,000~K the helium abundance could start
to decrease again, but the observed pattern
is hard to decipher. For these hot stars helium is depleted between a
factor of 10 and 100, without a clear trend with temperature, nor a
relation with the previously mentioned dichotomy on calculated masses.

\item We detect one EHB close binary candidate per cluster. Their RV
variations are quite small compared to typical sdB binary systems, but
the probability of their being due to random errors is negligible.
The candidate in \object{M\,80} is slightly cooler than typical EHB stars
(T$_\mathrm{eff}$=18\,100~K). None of them, nor any other EHB target,
show the MgIb triplet as signature of a cool companion, at variance
with the
only other sdB close binary discovered to date in a globular cluster
\citep{Moni08}. Therefore their companions are more likely compact
objects such as white dwarfs, or very low-mass main sequence stars.

\item The best estimate for the EHB close binary fraction in
\object{NGC\,5986} is $f$=25\%. This suggests that the fraction could be small, but
no value lower than 70\% can safely be excluded because of the small
observed sample. Nevertheless, the probability of a very high binary fraction
is low.

\item In \object{M\,80} EHB close binaries with period shorter than 5 days
are lacking with respect to what is observed among field sdB stars. The
best estimate is $f_\mathrm{P\leq 5days}$=12\%, and within a 90\%
confidence level $f_\mathrm{P\leq 5days}\leq$38\%. The fraction is not
very well determined, but even the lowest values found for field stars
are very improbable and can be ruled out. Results fairly agree with the
tiny $f$=4\% found in \object{NGC\,6752} \citep{Moni08}. \object{M\,80}
is the second cluster for which this behavior is observed. This
indicates that it is not just a peculiarity of \object{NGC\,6752}
but it should be quite a common feature, posing important new
constraints on models concerning sdB star formation.

\item Our results agree with the existence of an $f$-age relation
for sdB star populations, as proposed by \citet{Moni08} and theoretically
modeled by \citet{Han08} in the framework of binary scenarios for their
formation. Uncertainties in both predicted and observed
values prevent their direct comparison. Nevertheless, we find that
the $f$-age relation modeled by \citet{Han08} has some problem in
reproducing the details of the observations. A refined set of model
parameters could be needed, or the intervention of some
other effect able to affect $f$ and varying from cluster to cluster,
like for instance dynamical interactions in dense environment.

\end{itemize}

The low binary fraction among globular cluster EHB stars is consistent with 
the typical (low) binary fraction among globular cluster 
stars. After all, the high binary fraction of sdB stars might 
not be a peculiarity related to their nature of hot He burning
stars, but a simple consequence of the high binary fraction 
among field stars. If so, we still need to identify the cause
of sdB star formation.
Nevertheless, recent refined binary models can (at least
qualitatively) account for the lack of {\it close} EHB systems
in globular clusters, still retaining the hypothesis of a binary origin
\citep{Han08}. Thus, the "binary scenario" cannot be ruled out
by current observations, and we are still far from
a full understanding of the complete picture.


\begin{acknowledgements}
We want to thank the staff at the La Silla Paranal Observatory for their
support during our observations, and the referee for his comments and
suggestions, that improved the paper. GP Acknowledge support by MIUR
under the program PRIN2007 "Popolazioni Multiple in Ammassi GLobulari:
Censimento, Caratterizzazione, Origine".
\end{acknowledgements}


\bibliographystyle{aa}
\bibliography{biblioIII}

\end{document}